\message{UCBASIC.TEX TeX Macro Library}
\message{    A. Meiksin, 22-NOV-91     }




\def\HI{\hbox{H~$\scriptstyle\rm I\ $}}

\def\HeII{\hbox{He~$\scriptstyle\rm II\ $}}

\def\Lya{Ly$\alpha\ $}
\def\Lyb{Ly$\beta\ $}

\def\df{\leaders\hbox to 0.6em{\hss.}\hfill}


\def\section#1{\bigbreak\medskip\centerline{#1}\par\nobreak\medskip\markpage}

\def\subsection#1#2{\bigbreak\noindent{\bf#1\hskip 0.9em\relax#2}\par
   \nobreak\medskip\markpage}

\def\subsubsection#1#2{\medbreak\noindent{\sl#1\hskip 0.60em\relax#2}\par
   \nobreak\medskip\markpage}

\def\today{\advance\year by -1900 
   \number\month/\number\day/\number\year}
\def\yearmonthday{\number\year\space
   \ifcase\month\or January\or February\or March\or April\or May\or June\or
   July\or August\or September\or October\or November\or December\fi
   \space\number\day}

\newcount\num

\def\nextnum{\global\advance \num by 1 \number\num}
\def\nextitem{\leavevmode
   \hbox{\ifnum\num>8 \kern-0.43em\fi \nextnum.\kern0.60em}}
\def\bfnextitem{\leavevmode
   \hbox{\ifnum\num>8 \kern-0.43em\fi \bf\nextnum.\kern0.60em}}

\newcount\colnum

\def\nextcolnum{\global\advance \colnum by 1 \number\colnum}
\def\nextcolumn{\leavevmode
   \hbox{{\it \ifnum\colnum<9 \phantom{1}\fi Column \nextcolnum:}\kern0.60em}}

\newcount\fig

\def\nextfig{\global\advance \fig by 1 \number\fig}

\newcount\cap

\def\nextcap{\global\advance \cap by 1 \number\cap}

\newcount\letter

\def\nextlet{\global\advance \letter by 1
   \ifcase\letter\or A\or B\or C\or D\or E\or F\or G\or H\or I\or
   J\or K\or L\or M\or N\or O\or P\or Q\or R\or S\or T\or U\or V\or W\or X\or
   Y\or Z\fi}

\newdimen\bigindent \bigindent=3.5in
\def\letterhead{\hsize=6in\interlinepenalty=2000\parskip=6pt minus 3pt
  \pretolerance=750
  \def\topline##1{\hbox to\hsize{\hfil##1\hskip\rightskip}}
  \null
  \vskip-0.2in
  {\advance\rightskip by -0.75in
    \topline{Dept. of Astronomy \& Astrophysics}
    \topline{University of Chicago}
    \topline{5640 S. Ellis Ave.}
    \topline{Chicago, IL 60637}
    \topline{(312) 702-8203}
    \topline{BITNET:meiksin@oddjob.uchicago.edu}\par}
  \vskip30pt minus 15pt
  {\leftskip=\bigindent\yearmonthday\par}}
\def\myletterhead{\hsize=6in\interlinepenalty=2000\parskip=6pt minus 3pt
  \pretolerance=750
  \def\topline##1{\hbox to\hsize{\hfil##1\hskip\rightskip}}
  \null
  \vskip-0.2in
  {\advance\rightskip by -0.75in
    \topline{5020 S. Lake Shore Dr., Apt. 1715}
    \topline{Chicago, IL 60615}
    \topline{(312) 955-0152}\par}
  \vskip30pt minus 15pt
  {\leftskip=\bigindent\yearmonthday\par}}

\def\indentleft{\advance\leftskip by 50pt\interlinepenalty=750}
\def\inndentleft{\advance\leftskip by 78pt\interlinepenalty=750}
\def\narrower{\advance\leftskip by 0.42in\advance\rightskip by 0.42in
  \interlinepenalty=750}
\def\nnarrower{\advance\leftskip by 50pt\advance\rightskip by 45pt
  \interlinepenalty=750}

\def\checkbox{\nnarrower\parindent=0pt\itemitem{\vbox{\hrule height.7pt
  \hbox{\vrule width.7pt height6pt \kern6pt \vrule width.7pt}
  \hrule height.7pt}$\,$}}  


%
%
\newcount\index \index=100
\def\markpage{\advance\index by 1 \count\index=\pageno}
\def\begintableofcontents{\begingroup
  \index=100 \frenchspacing\interlinepenalty=750
  \parskip=0.1pt plus 1pt minus 0.1pt \parindent=0.3in
  \def\dfi{\advance\index by 1 \df\number\count\index}
  \def\in{\par\hskip-0.2in\indent \hangindent2\parindent \textindent}    
  \def\inin{\par\hskip0.32in\indent \hangindent3\parindent \textindent}
  \def\ininin{\par\hskip0.95in\indent \hangindent4\parindent \textindent}}



{\obeylines\gdef\startdisplay#1
  {\catcode`\^^M=5$$#1\halign\bgroup\indent##\hfil&&\qquad##\hfil\cr}}
\outer\def\enddisplay{\crcr\egroup$$}

\chardef\other=12
\def\ttverbatim{\begingroup \catcode`\\=\other \catcode`\{=\other
  \catcode`\}=\other \catcode`\$=\other \catcode`\&=\other
  \catcode`\#=\other \catcode`\%=\other \catcode`\~=\other
  \catcode`\_=\other \catcode`\^=\other
  \obeyspaces \obeylines \tt}
{\obeyspaces\gdef {\ }} 

\outer\def\begintt{$$\let\par=\endgraf \ttverbatim \parskip=0pt
  \catcode`\|=0 \rightskip=-5pc \ttfinish}
{\catcode`\|=0 \catcode`|\=\other 
  |obeylines 
  |gdef|ttfinish#1^^M#2\endtt{#1|vbox{#2}|endgroup$$}}

\catcode`\|=\active
{\obeylines\gdef|{\ttverbatim\spaceskip=\ttglue\let^^M=\ \let|=\endgroup}}

  \font\twentyfourrm=cmr10 scaled 2488
  \font\twentyfouri=cmmi10 scaled 2074   \font\twentyfoursy=cmsy10 scaled 2074
  \font\twentyrm=cmr10 scaled 2074      
  \font\twentyi=cmmi10 scaled 2074   \font\twentysy=cmsy10 scaled 2074
  \font\eighteenrm=cmr10 scaled 1728
  \font\eighteeni=cmmi10 scaled 1728 \font\eighteensy=cmsy10 scaled 1728
  \font\fourteenrm=cmr10 scaled 1440
  \font\fourteeni=cmmi10 scaled 1440 \font\fourteensy=cmsy10 scaled 1440
  \font\twelverm=cmr12

  \font\twelvei=cmmi12               \font\twelvesy=cmsy10 scaled 1200
  \font\elevenrm=cmr10 scaled 1095
    
  \font\eleveni=cmmi10 scaled 1095   \font\elevensy=cmsy10 scaled 1095
  \font\tenrm=cmr10

  \font\ninei=cmmi9                  \font\ninesy=cmsy9
  
  \font\seveni=cmmi7 \font\sevensy=cmsy7

\def\commonstuff{
  \parindent=0.42in       
  \def\skipline{\vskip\baselineskip}
  \hyphenpenalty=200\pretolerance=300\tolerance=600 
  \interlinepenalty=100\clubpenalty=500\widowpenalty=500
  \nonfrenchspacing\singlespace\rm}

\def\twelvepoint{
  \font\bf=cmbx12
  \font\it=cmti12
  \font\sl=cmsl12
  \font\tb=cmtt10 scaled 1200 
  \font\tt=cmtt8 scaled 1440
  \textfont0=\twelverm \scriptfont0=\tenrm     
    \scriptscriptfont0=\sevenrm                 
  \def\rm{\fam0 \twelverm}   
  \textfont1=\twelvei  \scriptfont1=\teni  
    \scriptscriptfont1=\seveni                  
  \def\mit{\fam1 } \def\oldstyle{\fam1 \twelvei}
  \textfont2=\twelvesy \scriptfont2=\tensy 
    \scriptscriptfont2=\sevensy                 
  \def\singlespace{\baselineskip=13.5pt\lineskiplimit=-5pt
    \lineskip=0pt
    \parskip=1.25pt plus 1.5pt minus 0.25pt}  
  \def\oneandahalfspace{\baselineskip=18pt\parskip=0pt plus 1pt}
  \def\doublespace{\baselineskip=24pt\parskip=0pt plus 0.5pt}
  \footline={\ifnum\pageno=1 \hfil
             \else\hss\twelverm-- \folio\ --\hss\fi} 
  \def\pagenumbers{\footline={\hss\twelverm-- \folio\ --\hss}}  
  \def\romanpagenumbers{\footline={\hss\twelverm-- \romannumeral\folio\ --\hss}}
  \commonstuff}

\def\apjtwelvepoint{
  \font\bf=cmbx12
  \font\it=cmti12
  \font\sl=cmsl12
  \font\tb=cmtt10 scaled 1200 
  \font\tt=cmtt8 scaled 1440
  \textfont0=\twelverm \scriptfont0=\tenrm     
    \scriptscriptfont0=\sevenrm                 
  \def\rm{\fam0 \twelverm}   
  \textfont1=\twelvei  \scriptfont1=\teni  
    \scriptscriptfont1=\seveni                  
  \def\mit{\fam1 } \def\oldstyle{\fam1 \twelvei}
  \textfont2=\twelvesy \scriptfont2=\tensy 
    \scriptscriptfont2=\sevensy                 
  \def\singlespace{\baselineskip=13.5pt\lineskiplimit=-5pt
    \lineskip=0pt
    \parskip=1.25pt plus 1.5pt minus 0.25pt}  
  \def\oneandahalfspace{\baselineskip=18pt\parskip=0pt plus 1pt}
  \def\doublespace{\baselineskip=24pt\parskip=0pt plus 0.5pt}
  \nopagenumbers
  \headline={\ifnum\pageno=1 \hfil
             \else\hss\twelverm-- \folio\ --\hss\fi} 
  \def\pagenumbers{\headline={\hss\twelverm-- \folio\ --\hss}}  
  \def\romanpagenumbers{\headline={\hss\twelverm-- \romannumeral\folio\ --\hss}}
  \commonstuff}

\def\tenpoint{
  \font\it=cmti10
  \font\sl=cmsl10
  \font\bf=cmb10
  \textfont0=\tenrm \scriptfont0=\sevenrm     
    \scriptscriptfont0=\fiverm                 
  \def\rm{\fam0 \tenrm}   
  \textfont1=\teni  \scriptfont1=\seveni  
    \scriptscriptfont1=\fivei                  
  \def\mit{\fam1 } \def\oldstyle{\fam1 \teni}
  \textfont2=\tensy \scriptfont2=\sevensy 
    \scriptscriptfont2=\fivesy                 
  \def\singlespace{\baselineskip=12pt\lineskiplimit=0pt
    \lineskip=-0.5mm       
    \parskip=2pt plus 1pt minus 1pt}  
  \footline={\ifnum\pageno=1 \hfil
             \else\hss\tenrm-- \folio\ --\hss\fi} 
  \def\oneandahalfspace{\baselineskip=18pt\parskip=0pt plus 1pt}
  \def\doublespace{\baselineskip=24pt\parskip=0pt plus 1 pt}
  \def\pagenumbers{\footline={\hss\tenrm-- \folio\ --\hss}}  
  \def\romanpagenumbers{\footline={\hss\tenrm-- \romannumeral\folio\ --\hss}}
  \commonstuff}

\def\elevenpoint{
  \font\it=cmti10 scaled 1095
  \font\sl=cmsl10 scaled 1095
  \font\bf=cmb10 scaled 1095 
  \font\tt=cmtt10 scaled 1095
  \textfont0=\elevenrm \scriptfont0=\tenrm     
    \scriptscriptfont0=\ninerm                 
  \def\rm{\fam0 \elevenrm}   
  \textfont1=\eleveni  \scriptfont1=\teni  
    \scriptscriptfont1=\ninei                  
  \def\mit{\fam1 } \def\oldstyle{\fam1 \eleveni}
  \textfont2=\elevensy \scriptfont2=\tensy 
    \scriptscriptfont2=\ninesy                 
  \def\singlespace{\baselineskip=13pt\lineskiplimit=-5pt
    \lineskip=0mm       
    \parskip=2pt plus 1pt minus 1pt}  
  \footline={\ifnum\pageno=1 \hfil
             \else\hss\elevenrm-- \folio\ --\hss\fi} 
  \def\oneandahalfspace{\baselineskip=19pt\parskip=0pt plus 1pt}
  \def\doublespace{\baselineskip=26pt\parskip=0pt plus 1 pt}
  \def\pagenumbers{\footline={\hss\elevenrm-- \folio\ --\hss}}  
  \def\romanpagenumbers{\footline={\hss\tenrm-- \romannumeral\folio\ --\hss}}
  \commonstuff}

\def\eighteenpoint{           
  \font\bf=cmb10 scaled 1728
  \font\it=cmti10 scaled 1728
  \font\sl=cmsl10 scaled 1728
  \font\tb=cmtt10 scaled 1728
  \font\tt=cmtt10 scaled 1728
  \textfont0=\eighteenrm \scriptfont0=\fourteenrm
    \scriptscriptfont0=\twelverm                 
  \def\rm{\fam0 \eighteenrm}   
  \textfont1=\eighteeni  \scriptfont1=\fourteeni  
    \scriptscriptfont1=\twelvei                  
  \def\mit{\fam1 } \def\oldstyle{\fam1 \eighteeni}
  \textfont2=\eighteensy \scriptfont2=\fourteensy 
    \scriptscriptfont2=\twelvesy                 
  \def\singlespace{\baselineskip=21pt\lineskiplimit=-5pt
    \lineskip=0pt
    \parskip=4pt plus 1pt minus 1pt}  
  \def\oneandahalfspace{\baselineskip=30pt\parskip=0pt plus 1pt}
  \def\doublespace{\baselineskip=40pt\parskip=0pt plus 1pt}
  \footline={\ifnum\pageno=1 \hfil
             \else\hss\eighteenrm-- \folio\ --\hss\fi} 
  \def\pagenumbers{\footline={\hss\eighteenrm-- \folio\ --\hss}}  
  \commonstuff}

\def\twentypoint{
  \font\bf=cmb10 scaled 2074
  \font\it=cmti10 scaled 2074
  \font\sl=cmsl10 scaled 2074
  \font\tb=cmtt10 scaled 2074
  \font\tt=cmtt10 scaled 2074
  \textfont0=\twentyrm \scriptfont0=\eighteenrm     
    \scriptscriptfont0=\fourteenrm                 
  \def\rm{\fam0 \twentyrm}   
  \textfont1=\twentyi  \scriptfont1=\eighteeni  
    \scriptscriptfont1=\fourteeni                  
  \def\mit{\fam1 } \def\oldstyle{\fam1 \twentyi}
  \textfont2=\twentysy \scriptfont2=\eighteensy 
    \scriptscriptfont2=\fourteensy                 
  \def\singlespace{\baselineskip=24pt\lineskiplimit=-5pt
    \lineskip=0pt
    \parskip=6pt plus 1pt minus 2pt}  
  \def\oneandahalfspace{\baselineskip=33pt\parskip=0pt plus 1pt}
  \def\doublespace{\baselineskip=44pt\parskip=0pt plus 0.5pt}
  \footline={\ifnum\pageno=1 \hfil
             \else\hss\twentyrm-- \folio\ --\hss\fi} 
  \def\pagenumbers{\footline={\hss\twentyrm-- \folio\ --\hss}}  
  \def\romanpagenumbers{\footline={\hss\twentyrm-- \romannumeral\folio\ --\hss}}
  \commonstuff}

\def\twentyfourpoint{
  \font\bf=cmb10 scaled 2488
  \font\it=cmti10 scaled 2488
  \font\sl=cmsl10 scaled 2488
  \font\tb=cmtt10 scaled 2488
  \font\tt=cmtt10 scaled 2488
  \textfont0=\twentyfourrm \scriptfont0=\twentyrm     
    \scriptscriptfont0=\eighteenrm                 
  \def\rm{\fam0 \twentyfourrm}   
  \textfont1=\twentyfouri  \scriptfont1=\twentyi  
    \scriptscriptfont1=\eighteeni                  
  \def\mit{\fam1 } \def\oldstyle{\fam1 \twentyfouri}
  \textfont2=\twentyfoursy \scriptfont2=\twentysy 
    \scriptscriptfont2=\eighteensy                 
  \def\singlespace{\baselineskip=28pt\lineskiplimit=-5pt
    \lineskip=0pt
    \parskip=5pt plus 1.5pt minus 1.5pt}  
  \def\oneandahalfspace{\baselineskip=42pt\parskip=0pt plus 1pt}
  \def\doublespace{\baselineskip=56pt\parskip=0pt plus 0.5pt}
  \footline={\ifnum\pageno=1 \hfil
             \else\hss\twentyfourrm-- \folio\ --\hss\fi} 
  \def\pagenumbers{\footline={\hss\twentyfourrm-- \folio\ --\hss}}  
  \def\romanpagenumbers{\footline={\hss\twentyfourrm-- \romannumeral\folio\ --\hss}}
  \commonstuff}

\twelvepoint

\def\page{\vfill\eject}

\def\ltsima{$\; \buildrel < \over \sim \;$}
\def\lsim{\lower.5ex\hbox{\ltsima}}
\def\gtsima{$\; \buildrel > \over \sim \;$}
\def\gsim{\lower.5ex\hbox{\gtsima}}

\def\lta{\lsim}
\def\HI{\hbox{H~$\scriptstyle\rm I\ $}}

\def\HeII{\hbox{He~$\scriptstyle\rm II\ $}}

\def\Lya{Ly$\alpha\ $}
\def\Lyb{Ly$\beta\ $}

\def\kmsmpc{\,{\rm km\,s^{-1}\,Mpc^{-1}}}
\def\cmm{\,{\rm cm^{-2}}}


\centerline{\bf Cosmological Filaments and Minivoids:}
\centerline{\bf The Origin of Intergalactic Absorption$^\#$}

\vskip 0.5truein
\centerline{Yu Zhang$^\dagger$$^*$, Avery Meiksin$^\ddagger$, 
            Peter Anninos$^\dagger$, \& Michael L. Norman$^\dagger$$^*$}
\vskip 0.1truein
\centerline{\it $^\dagger$Laboratory for Computational Astrophysics}
\centerline{\it           National Center for Supercomputing Applications}
\centerline{\it           University of Illinois at Urbana-Champaign}
\centerline{\it           405 N. Mathews Ave., Urbana, IL 61801}
\vskip 0.1truein
\centerline{\it$^\ddagger$Edwin Hubble Research Scientist}
\centerline{\it           University of Chicago}
\centerline{\it           Department of Astronomy \& Astrophysics}
\centerline{\it           5640 South Ellis Avenue}
\centerline{\it           Chicago, IL 60637}
\vskip 1.0truein

\vskip 0.2truein
\noindent{\bf
Soon after the first QSO was identified, Gunn \& Peterson$^{1}$ searched for
the expected characteristic absorption trough on the blueward side of \Lya in
the spectrum of the QSO due to an Intergalactic Medium (IGM).
They failed to find it, placing a constraint on the density of neutral
hydrogen in the IGM that was less than 1 part in $10^5$ of that residing in
galaxies, and concluded that the IGM must be highly ionized. Soon afterwards
absorption by \Lya was detected in the IGM; not by a diffuse component,
but by a clumpy component of intergalactic gas clouds,
the \Lya forest$^{2,3}$. The goal then became to search for `excess' absorption
beyond that expected from the forest. No clear absorption by a diffuse \HI
component, however, was ever detected$^{4,5}$.
The results of recent numerical hydrodynamics computations of the formation
of the \Lya forest appear to indicate that the division between a clumpy
component and a diffuse one may be inappropriate. We find that in a
CDM-dominated cosmology, no substantial diffuse medium should be expected.
Instead, the medium condenses into a network of complex structures 
that reveal themselves
as discrete absorption systems in the spectra of QSOs.  The lowest column
density lines arise from the fine structure in minivoids -- small
regions with densities below the cosmic mean.
These results suggest that the long sought for diffuse \HI Gunn-Peterson effect
may not exist.
} 

\vskip60pt
\noindent
$^\#$ {Submitted to \it Nature}

\noindent
$^*$ {\it Astronomy Department, University of Illinois at Urbana-Champaign}

\page

Recent ground and space measurements of intergalactic absorption in the
spectra of high redshift QSOs have considerably expanded our knowledge of
the structure of the IGM. Keck observations$^{6,7}$ have revealed that the
forest extends to \HI column densities as low as $N_{\rm HI}\sim10^{12}\cmm$,
while the {\it Hubble Space Telescope}$^8$ and the {\it Hopkins Ultraviolet
Telescope} (Davidsen et al., submitted to {\it Nature}),
have made the first detections of intergalactic \HeII.
The Keck data show no conspicuous absorption by a diffuse \HI component at
$z\sim3$. Absorption by diffuse \HeII in the IGM is expected to be more easily
detected than \HI because of its higher density, however it is unclear whether
the \HeII detections were of a diffuse component or of the \HeII content of the
\Lya forest$^{9}$. Indeed, Keck observations suggest the latter may account
for the measured opacity without invoking an unreasonably steep photoionizing
background radiation field.$^{9,10}$ In agreement with ground based \HI
observations, no diffuse component appears to be required.

We analyze the results of a recent cosmological hydrodynamics simulation of
the formation of the \Lya forest in a Cold Dark Matter (CDM) dominated
cosmology by Zhang, Anninos \& Norman$^{11}$ and derive their 
implications for \HI and \HeII
absorption. For much of our discussion, we have rescaled the ionization
fractions to the metagalactic radiation field recently determined by Haardt and
Madau$^{12}$ (HM) on the basis of QSO counts, including the re-emission from
the forest and Lyman-limit systems. The rescaling should have little impact on
the cloud properties other than on their ionization state, at least for the
column density range of interest here.  We show the distribution of gas density
in Figure 1. For the density threshold shown, the IGM appears as a
network of filaments, with a
typical coherence length of 1--2 Mpc and thickness 100--200 kpc.
These scales are comparable to the recent size estimates based on neighboring
QSO pairs$^{13,14}$. 
In between the filaments are underdense regions, or
cosmic `minivoids.'

To translate the \HI density into spectra, we lay down random lines of sight
through the box and determine the absorption features that would be observed
in a background QSO. We then identify absorption features with optical depths
exceeding 0.2, de-blend the features into individual lines, fit the lines
and extract the column densities and Doppler widths using a curve-of-growth
analysis, as
described by Zhang, Anninos, Norman, \& Meiksin (in preparation). 
The column density distribution
matches the observed distribution in shape remarkably well, as has been
found in other simulations for CDM (Hernquist {\it et al.} submitted to
{\it Astrophys. J}) and CDM $+$ $\Lambda^{15}$
cosmologies. The number of systems per unit column density varies as a power
law, $dN/dN_{\rm HI}\propto N_{\rm HI}^{-\beta}$, with $\beta\simeq1.5$ 
for $N_{\rm HI}<10^{14}\cmm$, in agreement with findings from the Keck$^7$.
There is an essential uncertainty in the \HI column densities, however,
resulting from the unknown ionization fractions. The neutral fraction is
proportional to $b_{\rm ion}\equiv(\Omega_bh^2_{50})^2/J_{912}$ (for clouds
optically thin at the \HI Lyman edge), where $J_{912}$ is the background
intensity at the \HI Lyman edge. This acts effectively as an `ionization bias'
relating the neutral hydrogen column density of an absorber to its total
hydrogen column density. Defining $b_{\rm ion}=1$ for $\Omega_bh_{50}^2=0.06$
and the HM value for the background intensity, we find that
rescaling the background radiation to $b_{\rm ion}=1.5$
matches the number density of $N_{\rm HI}>10^{13}\cmm$ systems measured by the
Keck$^7$ at $z=3$. We assume this value throughout. It is noteworthy that this
value would place the required $\Omega_b$ close to the nucleosynthesis upper
limit$^{16}$ of $\Omega_bh_{50}^2<0.08$
for the HM spectrum. If the metagalactic radiation field much exceeds the
estimated contribution from QSOs, or if the density of baryons is lower, as
suggested by some recent deuterium measurements in high redshift absorption
systems$^{17}$, then this cosmological model would appear to be in conflict
with the measured numbers of clouds in the \Lya forest. 

A wide range of local densities is responsible for the absorbers. In Figure 2
we show the average column density as a function of the local average density,
normalized by the average cosmological value. Those systems which are
optically thin at line-center arise in minivoids. In fact, we find that
at redshift $z=3$ more than 70\% of the baryons reside in overdense regions,
while the remainder, in underdense regions, produces most of the low column
density ($10^{11}-10^{13}$ cm$^{-2}$) systems.
These results suggest that the long
sought for diffuse component of the IGM may not exist, but that clumping is
ubiquitous, even in regions at the average cosmic density and below. The view
that QSOs dominate the background UV radiation field and that the \Lya forest
captures most of the baryons at $z>2$ is fully consistent with current
observations$^{18,12}$. This view differs dramatically from the prevailing
model of the IGM in which the \Lya forest is composed of clouds embedded in a
pervasive diffuse medium that survived the onset of structure formation at high
redshift and contained the majority of the baryons created during the Big Bang.

Most of the absorption in a QSO spectrum shortward of \Lya results from line-
blanketing, the stochastic overlapping of absorption features. Searches for the
Gunn-Peterson effect have been attempts to detect excess absorption above that
due to the forest$^{4,5}$. Since the forest extends to very low column
densities, however, the distinction between the two becomes somewhat semantic.
But while line-blanketing and optically thin `Gunn-Peterson' absorption are
related, they are not identical. The connection may be made concrete by
appealing to the effective opacity $\tau_{\rm eff}$, defined by
$\exp(-\tau_{\rm eff})=\langle\exp(-\tau_\nu)\rangle$. For
line-blanketing$^{19}$,
$$
\tau_{\rm eff}\equiv{{1+z}\over{\lambda_0}}\int{dW {{\partial^2N}\over
{\partial z\partial W}}W}, \eqno(1)
$$
for a cloud distribution of redshifts and rest-frame equivalent widths
$\partial^2N/\partial z\partial W$. Restricting the range to optically thin
clouds at line-center gives $\tau_{\rm eff}\rightarrow \tau_
{\rm GP}^{\rm thin}$, where $\tau_{\rm GP}^{\rm thin}$ is the Gunn-Peterson
opacity corresponding to the spatially averaged neutral hydrogen density
$\langle n_{\rm HI}^{\rm thin}\rangle$ of the optically thin systems$^{20,1}$,
$$
\tau_{\rm GP}^{\rm thin} = {{s_u\lambda_0}\over{H_0}}
{{\langle n_{\rm HI}^{\rm thin}\rangle}\over{(1+z)(1+2q_0z)^{1/2}}}, \eqno(2)
$$
where $s_u$ is the frequency integrated absorption cross-section$^{21}$.
According to Figure 2, clouds of unit optical depth correspond to regions near
the average cosmological baryon density. Thus, we may identify the \HI
Gunn-Peterson effect with the optically thin tail of the \Lya forest. By
contrast, the opacity resulting from the high column density systems measures
the product of the Doppler parameter of the clouds and their number per unit
line-of-sight, above a given column density, or line-center opacity, threshold.
This is because an optically thick cloud enters the flat part of the
curve-of-growth, for which the equivalent width is proportional to the Doppler
width and insensitive to the \HI density.
(Note, though, that the number of clouds above a fixed
opacity threshold is sensitive to the internal cloud \HI density.) The physical
distinction in the character of these two limits may be exploited to test two
different aspects of the nature of the clouds and their evolution.
In Figure 3a, we show the evolution of $\tau_{\rm eff}$ for optically thin
($0.2<\tau_0<1$ and $0.5<\tau_0<1$) and optically thick ($\tau_0>10$) systems.
We also show the contribution from all the clouds with $\tau_0 > 0.2$.
(Only clouds optically thin at the Lyman edge
are counted since we have not corrected for radiative transfer effects.) We
find good agreement with the expected amount of blanketing inferred from
observations over the limited redshift range currently available.

A quantity closely related to the effective opacity is ${\rm D_A}$, the flux
decrement in a QSO spectrum between \Lya and \Lyb$^{22}$,
$$
{\rm D_A} = {{5}\over{32}}{1\over{(1+z_Q)}}\int_{(27/32)(1+z_Q)-1}^{z_Q}
\left[1-\exp(-\tau_\nu)\right]dz, \eqno(3)
$$
where $z_Q$ is the redshift of the QSO. In Figure 3b, we show the evolution of
${\rm D_A}$ obtained from the simulation, including cuts based on
overdensity. The distinction between the optically thin and optically thick
absorbers is clear: the absorption due to the optically thin material,
occupying regions near the average cosmic density, behaves like the
Gunn-Peterson opacity for $\tau_{\rm GP}\ll1$. 
We also compute ${\rm D_B}$, the decrement between \Lyb
and the Lyman limit$^{22}$. We find ${\rm D_B/D_A} =$ (1.1, 1.2, 1.2, 3.3) for
$z =$ (2, 3, 4, 5), compared to the average measured ratios$^{23,24}$ of
1.5 at $z=3$ and 1.2 at $z=4$.

The recent determinations of the \HeII \Lya opacity, $\tau_{304}$, may be used
to constrain the shape of the background UV radiation field$^{8-10}$. The
\HeII opacity predicted, however, requires an extrapolation of the \HI column
density distribution to lower than measured values as well as knowledge of
the unmeasured \HeII Doppler parameters$^{9}$. These quantities are predicted
by the simulation, and so it is of interest to infer the constraint on the
radiation field the simulation implies. In Figure 4, we show the evolution of
$\tau_{304}$ for different values of the softness parameter
$S_L\equiv J_{912}/ J_{228}$, the ratio of intensities at the \HI and \HeII
Lyman edges. We find that in order to match the observed values, $S_L>60$ is
required, a value consistent with a QSO-dominated UV background, though for
a somewhat soft intrinsic QSO spectrum $(\alpha\gsim1.5)^9$.
A \HeII opacity of $\tau_{304}>2$ at $z=3$ would be difficult to reconcile
with a QSO dominated background. It would then be necessary to appeal to
additional sources of \HI ionization, like an early generation of stars or
decaying neutrinos$^{26}$.

We point out that substantial fluctuations in the
opacity are expected on small scales, as shown in Figure 1. These fluctuations
result from the large density fluctuations expected on small scales in a CDM
cosmology. While most of the \HI blanketing arises from systems in regions
near the average cosmological density, the \HeII blanketing is dominated by
the smaller scale systems in the underdense regions. The \HeII opacity thus
serves as a probe of the minivoids.

We conclude that the IGM may be highly clumped, with a negligible baryon
fraction residing in a diffuse component. It is for this reason that a diffuse
component may have evaded detection in the form of the Gunn-Peterson effect.
The canonical picture of the IGM in which the absorption clouds are embedded
in a diffuse background medium is not viable in a CDM dominated cosmology.
It is replaced by a continuous spectrum of clumpiness:\ from the high
overdensity halos which give rise to the
high column density absorption lines, to the mildly overdense filaments which
give rise to the intermediate column density absorbers, to the underdense
minivoids which are responsible for most of the low column density systems
that have just recently been discovered.

\page

\noindent {\bf Figure Captions}

\vskip 0.2truein
\noindent
{Fig. 1. ---} 
Distribution of the gas
density at $z=3$ from a numerical hydrodynamics
simulation of the \Lya forest.  The simulation adopted a CDM spectrum of
primordial density fluctuations, normalized to the second year COBE
observations,
a Hubble constant of $H_0=50\kmsmpc$, a comoving box size of 9.6 Mpc,
and baryonic density of $\Omega_b=0.06$ composed of
76\% hydrogen and 24\% helium. The region shown is 2.4 Mpc (proper) on
a side.
The isosurfaces represent baryons at ten times the mean cosmic density
(characteristic of typical filamentary structures) and
are color coded
to the gas temperature (dark blue = $3\times 10^4$ K, 
light blue = $3\times 10^5$ K).
We note that higher density contours trace out isolated spherical
structures typically found at the intersections of the filaments.
A single random slice through the cube is also shown, with the baryonic
overdensity represented by a rainbow--like color map changing from 
black (minimum) to red (maximum).
The \HeII mass fraction is shown with a wire mesh
in this same slice. Notice that there
is fine structure everywhere. To emphasize the fine structure in the
minivoids, we have rescaled the mass fraction in the
overdense regions by the gas overdensity wherever it
exceeds unity.

\vskip 0.2truein
\noindent
{Fig. 2. ---} 
Distribution of \HI column density of the clouds as a function of the local
baryon overdensity. 
The vertical solid line is the dividing 
line between the overdense and underdense regions.
The horizontal dashed line represents the column density 
$N_{\rm HI}=10^{16.9}$ $\cmm$
above which the \HI opacity at the Lyman edge
exceeds 0.5 and radiation transport effects become important.
Discrete absorption
systems exhibit the full range of densities, and arise even in regions that
are underdense. The clouds become optically thin in \Lya at line-center
for $N_{\rm HI}\lta10^{13}$ $\cmm$. Optically thin systems occur predominantly
in the underdense regions, although there is a smooth transition between
optically thin and thick absorbers.

\vskip 0.2truein
\noindent
{Fig. 3. ---} 
(a)\ Effective opacity $\tau_{\rm eff}$ for optically thin ($0.2<\tau_0<1$ and
$0.5<\tau_0<1$) and optically thick ($\tau_0>10$) \Lya clouds.
(A lower limit of $\tau_0>0.5$ for the thin systems is required to avoid
incompleteness in the Keck data.)
While the optically thin opacity traces the spatially averaged neutral \HI
density of the clouds, the opacity based on the optically thick systems
traces the Doppler width and number of the clouds per unit line-of-sight,
since the equivalent width of a saturated line is given by$^{21}$
$W\simeq(2b/c)\lambda_0(\log \tau_0)^{1/2}$, nearly independent of $\tau_0$,
the line center opacity.
These opacities test independent properties of the clouds, and both agree
closely with the measured values from Keck observations$^7$. 
We also show the full contribution of
all the lines with $\tau_0>0.2$ to the effective opacity. These values again
agree well with the estimate from the Keck data.
(b)\ \HI flux decrement ${\rm D_A}$ due to \HI
\Lya absorption. The contributions from several overdensity cuts are shown
separately, corresponding closely to cuts in \HI column density. The measured
values$^{23-25}$ compare well with the simulation results. 
We note that the  agreement for
$z>3$ may be improved by allowing for an additional contribution to the UV
background from QSOs obscured by dust in damped \Lya systems$^{27}$.
The thin solid line shows the contribution to ${\rm D_A}$ from the lines,
computed as ${\rm D_L}=1-\exp
(-\tau_{\rm eff})$. This indicates that
the decrement is due almost entirely to line-blanketing.
Also shown (dotted line) is the Gunn-Peterson opacity obtained by summing the
contribution to the absorption from underdense regions as in equation
(2), and expressed as a flux decrement ${\rm D_{GP}}$. The Gunn-Peterson value
correlates strongly with the contribution of the underdense material to
${\rm D_A}$ for $\tau_{\rm GP}\ll1$, demonstrating that the \HI in the
underdense regions is optically thin in \Lya. Accordingly, the absorption
arising from the underdense regions may be associated with the Gunn-Peterson
effect. Its small value reflects the small fraction of the baryons residing in
the underdense regions.

\vskip 0.2truein
\noindent
{Fig. 4. ---} 
\HeII \Lya opacity using the numerical
data rescaled by the Haardt \& Madau$^{12}$  radiation spectrum, with
$b_{\rm ion}=1.5$. We reduce the \HeII spectrum amplitude with respect to the
\HI by factors of 2, 5 and 10 over the HM spectrum, corresponding to
softness parameters of $S_L=60$, 200 and 400 respectively.
The cross is the $1\sigma$ result of 3.2, with an error of $(+\infty, -1.1)$,
at $z=3.2$ from the {\it HST FOC} observation of Jakobsen et al.$^{8}$  and the
filled square is their 90\% confidence lower limit of 1.7. The filled circle is from
the {\it ASTRO-2 HUT} observation of Davidsen et al. (submitted to {\it Nature})
of $1.00 \pm 0.07$ at an average redshift $<z>=2.4$.

\page

1. Gunn, J.~E. \& Peterson, B.~A. Astrophys. J. 142, 1633--1636 (1965).

2. Lynds, C.~R. Astrophys. J. 164, L73--L78 (1971).

3. Sargent, W.~L.~W., Young, P.~J., Boksenberg, A. \& Tytler, D. Astrophys. J.
Suppl. Ser. 42, 41--81 (1980).

4. Steidel, C.~C. \& Sargent, W.~L.~W. Astrophys. J. 318, L11-L13 (1987).

5. Giallongo, E., D'Odorico, S., Fontana, A., McMahon, R.~G., Savaglio, S.,
Cristiani, S., Molaro, P. \& Trevese, D. Astrophys. J. 425, L1--L4 (1994).

6. Tytler, D., Fan, X.-M., Burles, S., Cottrell, L., Davis, C., Kirkman, D.
\& Zuo, L. in {\it QSO Absorption Lines} (ed. Meylan, G.) (Springer, 1995).

7. Hu, E.~M., Kim, T.-S., Cowie, L.~L., Songaila, A. \& Rauch, M. Astron. J.
110, 1526--1543 (1995).

8. Jakobsen, P., Boksenberg, A., Deharveng, J.~M., Greenfield, P.,
Jedrzejewski, R. \& Paresce, F. Nature 370, 35--39 (1994).

9. Madau, M. \& Meiksin, A. Astrophys. J. 433, L53--L56 (1994).

10. Songaila, A., Hu, E.~M. \& Cowie, L.~L. Nature, L124--L126 (1995).

11. Zhang, Y., Anninos, P. \& Norman, M.~L. Astrophys. J. 453, L57--L60 (1995).

12. Haardt, F. \& Madau, P. Astrophys. J. (in press) (1996).

13. Dinshaw, N., Foltz, C.~B., Impey, C.~D., Weymann, R.~J. \& Morris, S.~L.
Nature 373, 223--225 (1995).

14. Smette, A., Surdej, J., Shaver, P.~A., Reimers, D., Wisotzki, L. \&
K\" ohler, T. Astr. \& Astrophys. (in press) (1996).

15. Cen, R., Miralda-Escude, J., Ostriker, J.~P. \& Rauch, M. Astrophys. J.
437, L9--L12 (1994).

16. Copi, C.~J., Schramm, D.~N. \& Turner, M.~S. Astrophys. J. 455, L95--L98 (1995).

17. Songaila, A., Cowie, L.~L., Hogan, C.~J., and Rugers, M. Nature, 368, 599-604 (1994).

18. Meiksin, A. \& Madau, P. Astrophys. J. 412, 34--55 (1993).

19. Press, W.~H., Rybicki, G.~B. \& Schneider, D.~P. Astrophys. J. 414, 64--81
(1993).

20. Field, G.~B. Astrophys. J. 129, 536--550 (1959).

21. Spitzer, L. {\it Physical Processes in the Interstellar Medium}
(John Wiley \& Sons, 1978).

22. Oke, J.~B. \& Korycansky, D.~G. Astrophys. J. 255, 11--19 (1982).

23. Steidel, C.~L. \& Sargent, W.~L.~W. Astrophys. J. 313, 171--184 (1987).

24. Schneider, D.~P., Schmidt, M. \& Gunn, J.~E. Astron. J. 101, 2004--2016 (1991).

25. Kennefick, J.~D., deCarvalho, R.~R., Djorgovski, S.~G., Wilber, M.~M.,
Dickson, E.~S., Weir, N., Fayyad, U. \& Roden, J. Astron. J. 110, 78--86 (1995).

26. Sciama, D.~W. {\it Modern Cosmology and the Dark Matter Problem}
(Cambridge University Press, 1993).

27. Fall, S.~M. \& Pei, Y.~C. Astrophys. J. 402, 479--492 (1993).

\bigskip

ACKNOWLEDEGMENT.  We are very grateful to John Shalf for his help
in producing Figure 1.
This work is supported in part by NSF under the auspices of
the Grand Challenge Cosmology Consortium (GC3).  
A.M. is grateful to the NCSA at UIUC for
its hospitality where part of this work was conducted and to the W. Gaertner
Fund at the University of Chicago for support.
The calculations were performed on the Convex C3880 system at the National
Center for Supercomputing Applications, University of Illinois at Urbana-
Champaign.

\end
\message{UCBASIC.TEX TeX Macro Library}
\message{    A. Meiksin, 22-NOV-91     }




\def\HI{\hbox{H~$\scriptstyle\rm I\ $}}

\def\HeII{\hbox{He~$\scriptstyle\rm II\ $}}

\def\Lya{Ly$\alpha\ $}
\def\Lyb{Ly$\beta\ $}

\def\df{\leaders\hbox to 0.6em{\hss.}\hfill}


\def\section#1{\bigbreak\medskip\centerline{#1}\par\nobreak\medskip\markpage}

\def\subsection#1#2{\bigbreak\noindent{\bf#1\hskip 0.9em\relax#2}\par
   \nobreak\medskip\markpage}

\def\subsubsection#1#2{\medbreak\noindent{\sl#1\hskip 0.60em\relax#2}\par
   \nobreak\medskip\markpage}

\def\today{\advance\year by -1900 
   \number\month/\number\day/\number\year}
\def\yearmonthday{\number\year\space
   \ifcase\month\or January\or February\or March\or April\or May\or June\or
   July\or August\or September\or October\or November\or December\fi
   \space\number\day}

\newcount\num

\def\nextnum{\global\advance \num by 1 \number\num}
\def\nextitem{\leavevmode
   \hbox{\ifnum\num>8 \kern-0.43em\fi \nextnum.\kern0.60em}}
\def\bfnextitem{\leavevmode
   \hbox{\ifnum\num>8 \kern-0.43em\fi \bf\nextnum.\kern0.60em}}

\newcount\colnum

\def\nextcolnum{\global\advance \colnum by 1 \number\colnum}
\def\nextcolumn{\leavevmode
   \hbox{{\it \ifnum\colnum<9 \phantom{1}\fi Column \nextcolnum:}\kern0.60em}}

\newcount\fig

\def\nextfig{\global\advance \fig by 1 \number\fig}

\newcount\cap

\def\nextcap{\global\advance \cap by 1 \number\cap}

\newcount\letter

\def\nextlet{\global\advance \letter by 1
   \ifcase\letter\or A\or B\or C\or D\or E\or F\or G\or H\or I\or
   J\or K\or L\or M\or N\or O\or P\or Q\or R\or S\or T\or U\or V\or W\or X\or
   Y\or Z\fi}

\newdimen\bigindent \bigindent=3.5in
\def\letterhead{\hsize=6in\interlinepenalty=2000\parskip=6pt minus 3pt
  \pretolerance=750
  \def\topline##1{\hbox to\hsize{\hfil##1\hskip\rightskip}}
  \null
  \vskip-0.2in
  {\advance\rightskip by -0.75in
    \topline{Dept. of Astronomy \& Astrophysics}
    \topline{University of Chicago}
    \topline{5640 S. Ellis Ave.}
    \topline{Chicago, IL 60637}
    \topline{(312) 702-8203}
    \topline{BITNET:meiksin@oddjob.uchicago.edu}\par}
  \vskip30pt minus 15pt
  {\leftskip=\bigindent\yearmonthday\par}}
\def\myletterhead{\hsize=6in\interlinepenalty=2000\parskip=6pt minus 3pt
  \pretolerance=750
  \def\topline##1{\hbox to\hsize{\hfil##1\hskip\rightskip}}
  \null
  \vskip-0.2in
  {\advance\rightskip by -0.75in
    \topline{5020 S. Lake Shore Dr., Apt. 1715}
    \topline{Chicago, IL 60615}
    \topline{(312) 955-0152}\par}
  \vskip30pt minus 15pt
  {\leftskip=\bigindent\yearmonthday\par}}

\def\indentleft{\advance\leftskip by 50pt\interlinepenalty=750}
\def\inndentleft{\advance\leftskip by 78pt\interlinepenalty=750}
\def\narrower{\advance\leftskip by 0.42in\advance\rightskip by 0.42in
  \interlinepenalty=750}
\def\nnarrower{\advance\leftskip by 50pt\advance\rightskip by 45pt
  \interlinepenalty=750}

\def\checkbox{\nnarrower\parindent=0pt\itemitem{\vbox{\hrule height.7pt
  \hbox{\vrule width.7pt height6pt \kern6pt \vrule width.7pt}
  \hrule height.7pt}$\,$}}  


%
%
\newcount\index \index=100
\def\markpage{\advance\index by 1 \count\index=\pageno}
\def\begintableofcontents{\begingroup
  \index=100 \frenchspacing\interlinepenalty=750
  \parskip=0.1pt plus 1pt minus 0.1pt \parindent=0.3in
  \def\dfi{\advance\index by 1 \df\number\count\index}
  \def\in{\par\hskip-0.2in\indent \hangindent2\parindent \textindent}    
  \def\inin{\par\hskip0.32in\indent \hangindent3\parindent \textindent}
  \def\ininin{\par\hskip0.95in\indent \hangindent4\parindent \textindent}}



{\obeylines\gdef\startdisplay#1
  {\catcode`\^^M=5$$#1\halign\bgroup\indent##\hfil&&\qquad##\hfil\cr}}
\outer\def\enddisplay{\crcr\egroup$$}

\chardef\other=12
\def\ttverbatim{\begingroup \catcode`\\=\other \catcode`\{=\other
  \catcode`\}=\other \catcode`\$=\other \catcode`\&=\other
  \catcode`\#=\other \catcode`\%=\other \catcode`\~=\other
  \catcode`\_=\other \catcode`\^=\other
  \obeyspaces \obeylines \tt}
{\obeyspaces\gdef {\ }} 

\outer\def\begintt{$$\let\par=\endgraf \ttverbatim \parskip=0pt
  \catcode`\|=0 \rightskip=-5pc \ttfinish}
{\catcode`\|=0 \catcode`|\=\other 
  |obeylines 
  |gdef|ttfinish#1^^M#2\endtt{#1|vbox{#2}|endgroup$$}}

\catcode`\|=\active
{\obeylines\gdef|{\ttverbatim\spaceskip=\ttglue\let^^M=\ \let|=\endgroup}}

  \font\twentyfourrm=cmr10 scaled 2488
  \font\twentyfouri=cmmi10 scaled 2074   \font\twentyfoursy=cmsy10 scaled 2074
  \font\twentyrm=cmr10 scaled 2074      
  \font\twentyi=cmmi10 scaled 2074   \font\twentysy=cmsy10 scaled 2074
  \font\eighteenrm=cmr10 scaled 1728
  \font\eighteeni=cmmi10 scaled 1728 \font\eighteensy=cmsy10 scaled 1728
  \font\fourteenrm=cmr10 scaled 1440
  \font\fourteeni=cmmi10 scaled 1440 \font\fourteensy=cmsy10 scaled 1440
  \font\twelverm=cmr12

  \font\twelvei=cmmi12               \font\twelvesy=cmsy10 scaled 1200
  \font\elevenrm=cmr10 scaled 1095
    
  \font\eleveni=cmmi10 scaled 1095   \font\elevensy=cmsy10 scaled 1095
  \font\tenrm=cmr10

  \font\ninei=cmmi9                  \font\ninesy=cmsy9
  
  \font\seveni=cmmi7 \font\sevensy=cmsy7

\def\commonstuff{
  \parindent=0.42in       
  \def\skipline{\vskip\baselineskip}
  \hyphenpenalty=200\pretolerance=300\tolerance=600 
  \interlinepenalty=100\clubpenalty=500\widowpenalty=500
  \nonfrenchspacing\singlespace\rm}

\def\twelvepoint{
  \font\bf=cmbx12
  \font\it=cmti12
  \font\sl=cmsl12
  \font\tb=cmtt10 scaled 1200 
  \font\tt=cmtt8 scaled 1440
  \textfont0=\twelverm \scriptfont0=\tenrm     
    \scriptscriptfont0=\sevenrm                 
  \def\rm{\fam0 \twelverm}   
  \textfont1=\twelvei  \scriptfont1=\teni  
    \scriptscriptfont1=\seveni                  
  \def\mit{\fam1 } \def\oldstyle{\fam1 \twelvei}
  \textfont2=\twelvesy \scriptfont2=\tensy 
    \scriptscriptfont2=\sevensy                 
  \def\singlespace{\baselineskip=13.5pt\lineskiplimit=-5pt
    \lineskip=0pt
    \parskip=1.25pt plus 1.5pt minus 0.25pt}  
  \def\oneandahalfspace{\baselineskip=18pt\parskip=0pt plus 1pt}
  \def\doublespace{\baselineskip=24pt\parskip=0pt plus 0.5pt}
  \footline={\ifnum\pageno=1 \hfil
             \else\hss\twelverm-- \folio\ --\hss\fi} 
  \def\pagenumbers{\footline={\hss\twelverm-- \folio\ --\hss}}  
  \def\romanpagenumbers{\footline={\hss\twelverm-- \romannumeral\folio\ --\hss}}
  \commonstuff}

\def\apjtwelvepoint{
  \font\bf=cmbx12
  \font\it=cmti12
  \font\sl=cmsl12
  \font\tb=cmtt10 scaled 1200 
  \font\tt=cmtt8 scaled 1440
  \textfont0=\twelverm \scriptfont0=\tenrm     
    \scriptscriptfont0=\sevenrm                 
  \def\rm{\fam0 \twelverm}   
  \textfont1=\twelvei  \scriptfont1=\teni  
    \scriptscriptfont1=\seveni                  
  \def\mit{\fam1 } \def\oldstyle{\fam1 \twelvei}
  \textfont2=\twelvesy \scriptfont2=\tensy 
    \scriptscriptfont2=\sevensy                 
  \def\singlespace{\baselineskip=13.5pt\lineskiplimit=-5pt
    \lineskip=0pt
    \parskip=1.25pt plus 1.5pt minus 0.25pt}  
  \def\oneandahalfspace{\baselineskip=18pt\parskip=0pt plus 1pt}
  \def\doublespace{\baselineskip=24pt\parskip=0pt plus 0.5pt}
  \nopagenumbers
  \headline={\ifnum\pageno=1 \hfil
             \else\hss\twelverm-- \folio\ --\hss\fi} 
  \def\pagenumbers{\headline={\hss\twelverm-- \folio\ --\hss}}  
  \def\romanpagenumbers{\headline={\hss\twelverm-- \romannumeral\folio\ --\hss}}
  \commonstuff}

\def\tenpoint{
  \font\it=cmti10
  \font\sl=cmsl10
  \font\bf=cmb10
  \textfont0=\tenrm \scriptfont0=\sevenrm     
    \scriptscriptfont0=\fiverm                 
  \def\rm{\fam0 \tenrm}   
  \textfont1=\teni  \scriptfont1=\seveni  
    \scriptscriptfont1=\fivei                  
  \def\mit{\fam1 } \def\oldstyle{\fam1 \teni}
  \textfont2=\tensy \scriptfont2=\sevensy 
    \scriptscriptfont2=\fivesy                 
  \def\singlespace{\baselineskip=12pt\lineskiplimit=0pt
    \lineskip=-0.5mm       
    \parskip=2pt plus 1pt minus 1pt}  
  \footline={\ifnum\pageno=1 \hfil
             \else\hss\tenrm-- \folio\ --\hss\fi} 
  \def\oneandahalfspace{\baselineskip=18pt\parskip=0pt plus 1pt}
  \def\doublespace{\baselineskip=24pt\parskip=0pt plus 1 pt}
  \def\pagenumbers{\footline={\hss\tenrm-- \folio\ --\hss}}  
  \def\romanpagenumbers{\footline={\hss\tenrm-- \romannumeral\folio\ --\hss}}
  \commonstuff}

\def\elevenpoint{
  \font\it=cmti10 scaled 1095
  \font\sl=cmsl10 scaled 1095
  \font\bf=cmb10 scaled 1095 
  \font\tt=cmtt10 scaled 1095
  \textfont0=\elevenrm \scriptfont0=\tenrm     
    \scriptscriptfont0=\ninerm                 
  \def\rm{\fam0 \elevenrm}   
  \textfont1=\eleveni  \scriptfont1=\teni  
    \scriptscriptfont1=\ninei                  
  \def\mit{\fam1 } \def\oldstyle{\fam1 \eleveni}
  \textfont2=\elevensy \scriptfont2=\tensy 
    \scriptscriptfont2=\ninesy                 
  \def\singlespace{\baselineskip=13pt\lineskiplimit=-5pt
    \lineskip=0mm       
    \parskip=2pt plus 1pt minus 1pt}  
  \footline={\ifnum\pageno=1 \hfil
             \else\hss\elevenrm-- \folio\ --\hss\fi} 
  \def\oneandahalfspace{\baselineskip=19pt\parskip=0pt plus 1pt}
  \def\doublespace{\baselineskip=26pt\parskip=0pt plus 1 pt}
  \def\pagenumbers{\footline={\hss\elevenrm-- \folio\ --\hss}}  
  \def\romanpagenumbers{\footline={\hss\tenrm-- \romannumeral\folio\ --\hss}}
  \commonstuff}

\def\eighteenpoint{           
  \font\bf=cmb10 scaled 1728
  \font\it=cmti10 scaled 1728
  \font\sl=cmsl10 scaled 1728
  \font\tb=cmtt10 scaled 1728
  \font\tt=cmtt10 scaled 1728
  \textfont0=\eighteenrm \scriptfont0=\fourteenrm
    \scriptscriptfont0=\twelverm                 
  \def\rm{\fam0 \eighteenrm}   
  \textfont1=\eighteeni  \scriptfont1=\fourteeni  
    \scriptscriptfont1=\twelvei                  
  \def\mit{\fam1 } \def\oldstyle{\fam1 \eighteeni}
  \textfont2=\eighteensy \scriptfont2=\fourteensy 
    \scriptscriptfont2=\twelvesy                 
  \def\singlespace{\baselineskip=21pt\lineskiplimit=-5pt
    \lineskip=0pt
    \parskip=4pt plus 1pt minus 1pt}  
  \def\oneandahalfspace{\baselineskip=30pt\parskip=0pt plus 1pt}
  \def\doublespace{\baselineskip=40pt\parskip=0pt plus 1pt}
  \footline={\ifnum\pageno=1 \hfil
             \else\hss\eighteenrm-- \folio\ --\hss\fi} 
  \def\pagenumbers{\footline={\hss\eighteenrm-- \folio\ --\hss}}  
  \commonstuff}

\def\twentypoint{
  \font\bf=cmb10 scaled 2074
  \font\it=cmti10 scaled 2074
  \font\sl=cmsl10 scaled 2074
  \font\tb=cmtt10 scaled 2074
  \font\tt=cmtt10 scaled 2074
  \textfont0=\twentyrm \scriptfont0=\eighteenrm     
    \scriptscriptfont0=\fourteenrm                 
  \def\rm{\fam0 \twentyrm}   
  \textfont1=\twentyi  \scriptfont1=\eighteeni  
    \scriptscriptfont1=\fourteeni                  
  \def\mit{\fam1 } \def\oldstyle{\fam1 \twentyi}
  \textfont2=\twentysy \scriptfont2=\eighteensy 
    \scriptscriptfont2=\fourteensy                 
  \def\singlespace{\baselineskip=24pt\lineskiplimit=-5pt
    \lineskip=0pt
    \parskip=6pt plus 1pt minus 2pt}  
  \def\oneandahalfspace{\baselineskip=33pt\parskip=0pt plus 1pt}
  \def\doublespace{\baselineskip=44pt\parskip=0pt plus 0.5pt}
  \footline={\ifnum\pageno=1 \hfil
             \else\hss\twentyrm-- \folio\ --\hss\fi} 
  \def\pagenumbers{\footline={\hss\twentyrm-- \folio\ --\hss}}  
  \def\romanpagenumbers{\footline={\hss\twentyrm-- \romannumeral\folio\ --\hss}}
  \commonstuff}

\def\twentyfourpoint{
  \font\bf=cmb10 scaled 2488
  \font\it=cmti10 scaled 2488
  \font\sl=cmsl10 scaled 2488
  \font\tb=cmtt10 scaled 2488
  \font\tt=cmtt10 scaled 2488
  \textfont0=\twentyfourrm \scriptfont0=\twentyrm     
    \scriptscriptfont0=\eighteenrm                 
  \def\rm{\fam0 \twentyfourrm}   
  \textfont1=\twentyfouri  \scriptfont1=\twentyi  
    \scriptscriptfont1=\eighteeni                  
  \def\mit{\fam1 } \def\oldstyle{\fam1 \twentyfouri}
  \textfont2=\twentyfoursy \scriptfont2=\twentysy 
    \scriptscriptfont2=\eighteensy                 
  \def\singlespace{\baselineskip=28pt\lineskiplimit=-5pt
    \lineskip=0pt
    \parskip=5pt plus 1.5pt minus 1.5pt}  
  \def\oneandahalfspace{\baselineskip=42pt\parskip=0pt plus 1pt}
  \def\doublespace{\baselineskip=56pt\parskip=0pt plus 0.5pt}
  \footline={\ifnum\pageno=1 \hfil
             \else\hss\twentyfourrm-- \folio\ --\hss\fi} 
  \def\pagenumbers{\footline={\hss\twentyfourrm-- \folio\ --\hss}}  
  \def\romanpagenumbers{\footline={\hss\twentyfourrm-- \romannumeral\folio\ --\hss}}
  \commonstuff}
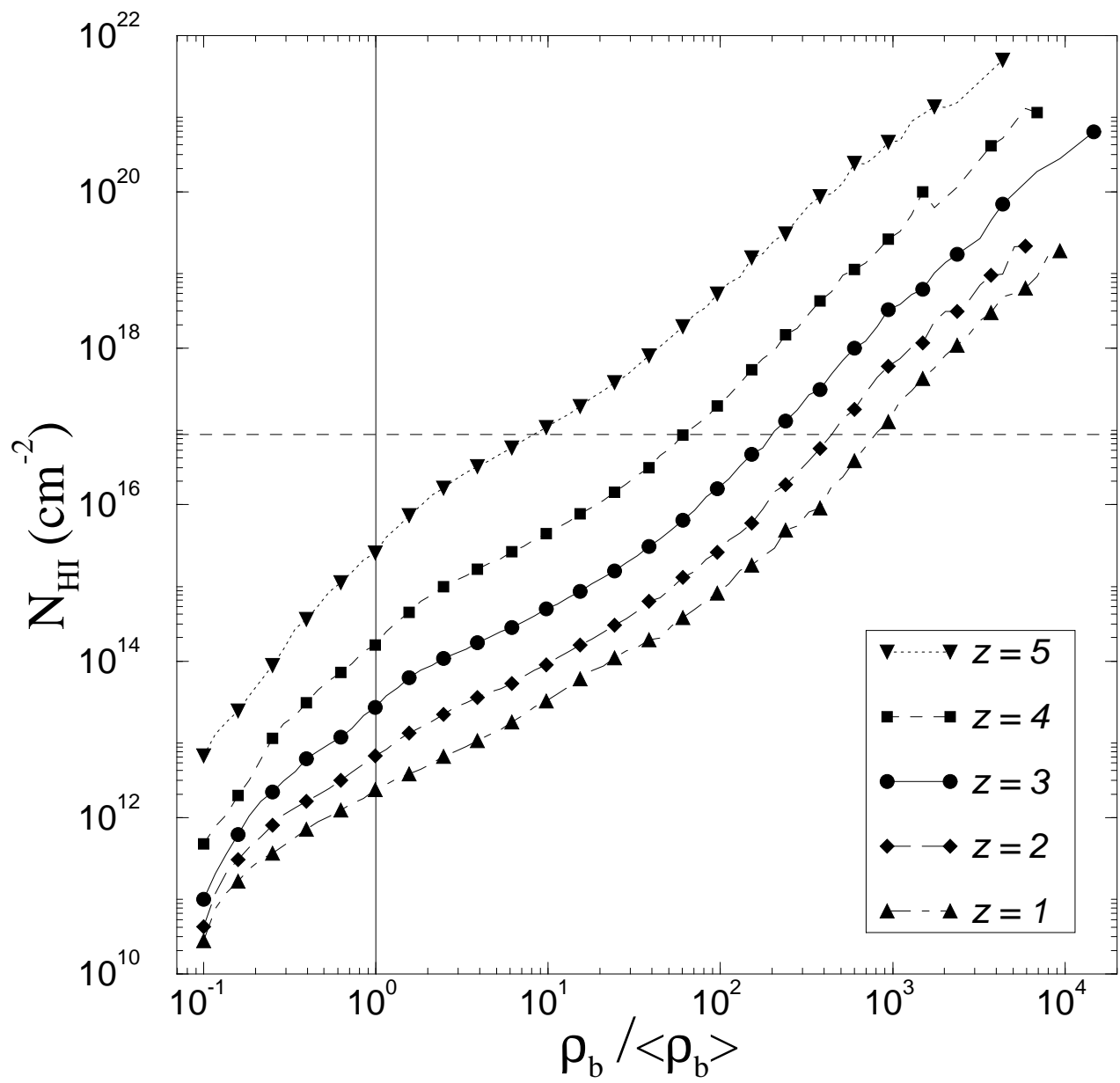
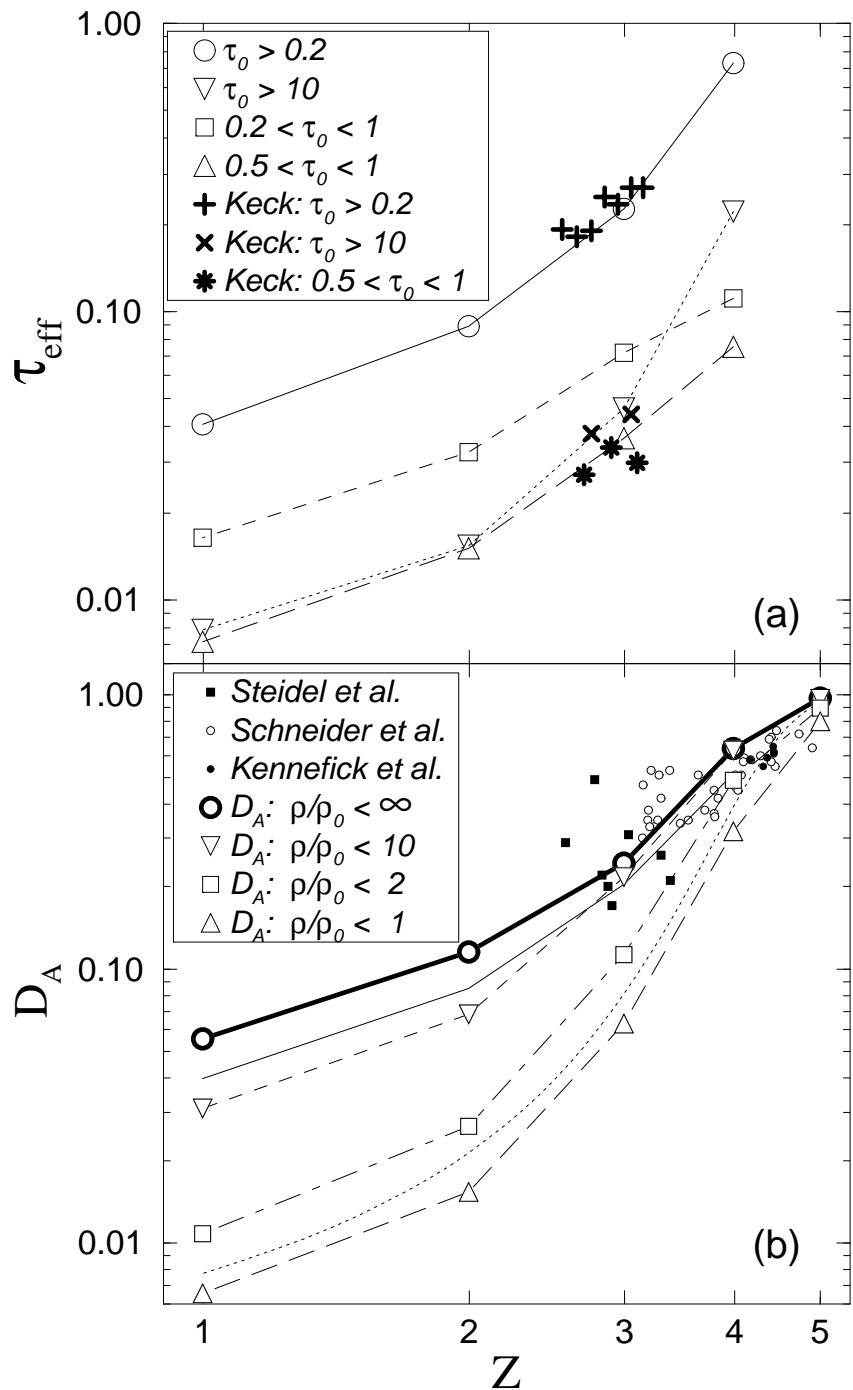
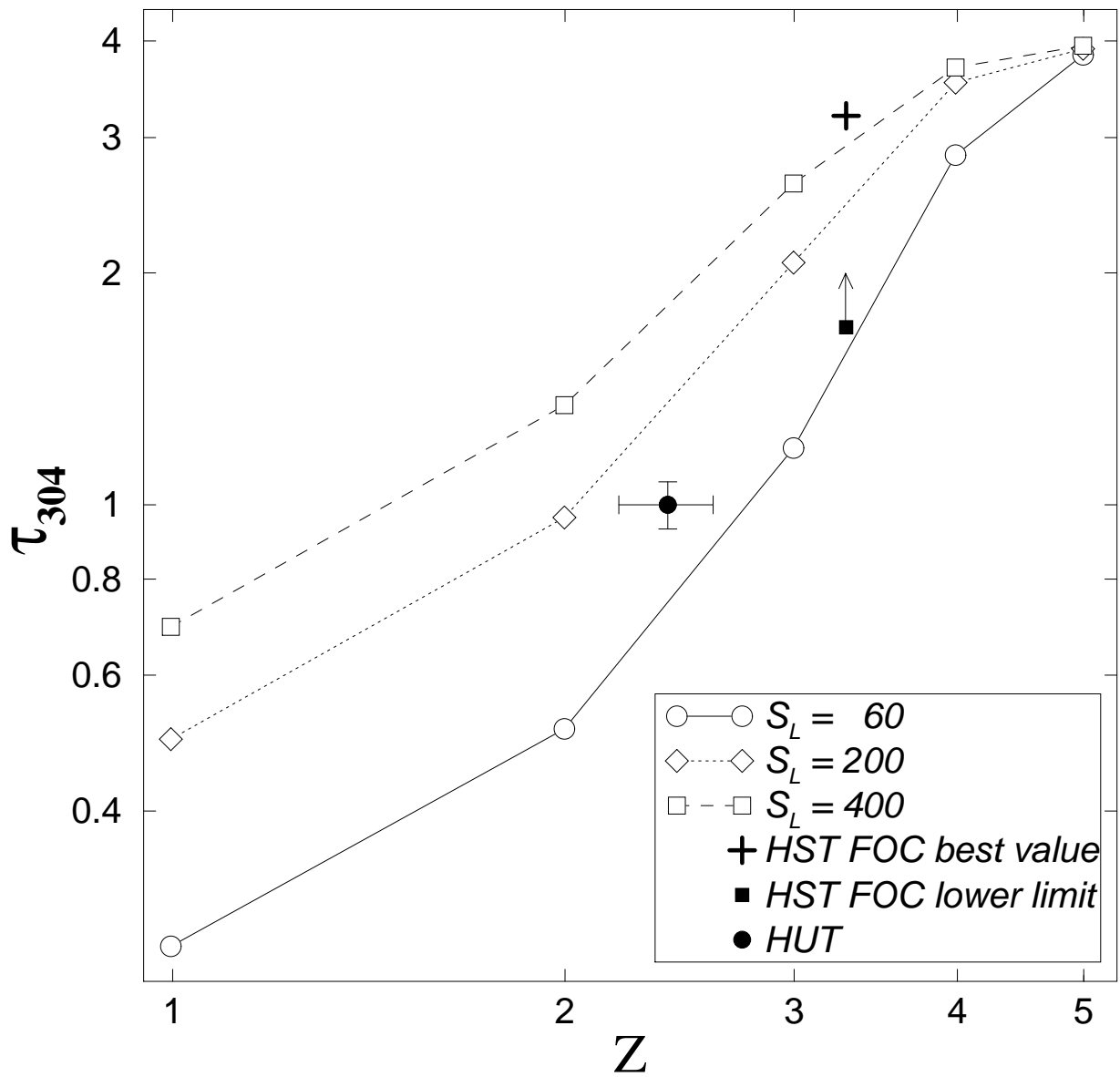
